\documentclass[preprint,aps,amsmath,superscriptaddress,nofootinbib]{revtex4}

\usepackage{dcolumn}
\usepackage{bm}
\usepackage{epsfig}
\usepackage{graphicx}
\usepackage{amsmath}
\usepackage{amssymb}
\usepackage{mathrsfs}
\usepackage{color}
\usepackage[usenames,dvipsnames]{xcolor}
\usepackage{hyperref}
\usepackage{braket}
\usepackage[caption=false]{subfig}
\usepackage{multirow}
\usepackage{array}
\usepackage{slashed}
\usepackage{rotating}

\usepackage{amsfonts}

\def\be{\begin{equation}}
\def\ee{\end{equation}}
\def\bea{\begin{eqnarray}}
\def\eea{\end{eqnarray}}

\newcommand{\pv}{{\bf p}}
\newcommand{\Pv}{{\bf P}}
\newcommand{\qv}{{\bf q}}

\newcommand{\kv}{{\bf k}}
\newcommand{\bv}{{\bf b}}

\newcommand{\zb}{\bar{z}}
\newcommand{\zt}{\tilde{z}}
\newcommand{\aem}{\alpha_{\rm em}}
\newcommand{\as}{\alpha_s}
\newcommand{\Pp}{{\bf P}_\perp}
\newcommand{\Otsosing}{\braket{\mathcal{O}^{J/\psi}(^3S_1^{[1]})}}
\newcommand{\Otsooct}{\braket{\mathcal{O}^{J/\psi}(^3S_1^{[8]})}}
\newcommand{\Oosz}{\braket{\mathcal{O}^{J/\psi}(^1S_0^{[8]})}}
\newcommand{\Otpz}{\braket{\mathcal{O}^{J/\psi}(^3P_0^{[8]})}}

\def\slashchar#1{\setbox0=\hbox{$#1$}           
   \dimen0=\wd0                                 
   \setbox1=\hbox{/} \dimen1=\wd1               
   \ifdim\dimen0>\dimen1                        
      \rlap{\hbox to \dimen0{\hfil/\hfil}}      
      #1                                        
   \else                                        
      \rlap{\hbox to \dimen1{\hfil$#1$\hfil}}   
      /                                         
   \fi}

\begin{document}

\title{Polarized $J/\psi$ production in semi-inclusive DIS at large $Q^2$: Comparing quark fragmentation and photon-gluon fusion}



\author{Marston Copeland}
\email{paul.copeland@duke.edu}
\affiliation{Department of Physics, Duke University, Durham, North Carolina\ 27708, USA\\}

\author{Sean Fleming}
\email{spf@email.arizona.edu}
\affiliation{Department of Physics, University of Arizona, Tucson, Arizona\ 85721, USA\\}

\author{Rohit Gupta}
\email{rohitkgupta@arizona.edu}
\affiliation{Department of Physics, University of Arizona, Tucson, Arizona\ 85721, USA\\}

\author{Reed Hodges}
\email{reed.hodges@duke.edu}
\affiliation{Department of Physics, Duke University, Durham, North Carolina\ 27708, USA\\}

\author{Thomas Mehen}
\email{mehen@duke.edu}
\affiliation{Department of Physics, Duke University, Durham, North Carolina\ 27708, USA\\}

\begin{abstract} 
We compare the relative importance of different mechanisms for polarized $J/\psi$ production in semi-inclusive deep inelastic scattering processes at large $Q^2$. The transverse momentum dependent (TMD) factorization framework and nonrelativistic quantum chromodynamics are used to study the leading contributions from light quark fragmentation to polarized $J/\psi$, and compared to direct production via photon-gluon fusion, which can proceed through color-singlet as well as color-octet mechanisms. We identify kinematic regimes where light quark fragmentation dominates, allowing for the extraction of the $^3S_1^{[8]}$ matrix element, as well as regimes where photon gluon fusion dominates, suggesting that the gluon TMD parton distribution function can be probed.

\end{abstract}

\maketitle

\section{Introduction}

The Electron-Ion Collider (EIC) \cite{Accardi:2012qut, AbdulKhalek:2021gbh} is scheduled to begin operations early in the next decade, and is expected to ``revolutionize our understanding of the inner workings of the nucleus by providing us with detailed knowledge of the internal structure of the proton...''\cite{yeck:cracow} One facet of understanding the structure of nucleons is the precision extraction of transverse momentum dependent (TMD) distributions, and the production of polarized $J/\psi$ offers a new and relatively unexplored method for studying these distributions. A thorough theoretical understanding of the various mechanisms responsible for polarized $J/\psi$ production is advantageous as it allows us to determine the kinematic regimes where the TMD distribution of partons (TMDPDFs) in the proton can be cleanly extracted. In particular, it is important to isolate regions of phase space where either the quark or gluon TMDPDF dominate so each can be measured independently. 

In this paper, we use an effective field theory (EFT) approach by combining the transverse momentum dependent
factorization framework and nonrelativistic QCD (NRQCD)~\cite{Bodwin:1994jh,Luke:1999kz,Brambilla:1999xf} to systematically study the most important channels for polarized $J/\psi$ production. Using the power counting of the EFT we conclude that there are leading contributions both from light quark fragmentation and 
photon-gluon fusion. In the limit of large center-of-mass energy, fragmentation is the dominant mechanism for inclusive production of any identified hadron in collisions. In the specific case of $J/\psi$ production via fragmentation, NRQCD power counting tells us that light quark fragmentation to a charm anticharm ($c\bar{c}$) pair in a color-octet ${}^3S_1$ configuration (the so-called color-octet mechanism) is the leading order process. No machine, however, has infinite energy so it is important to consider corrections to fragmentation. These are called direct production contributions. Again using the combined power counting of TMD factorization and NRQCD two important direct production processes which compete with fragmentation can be identified: photon-gluon fusion to either a color-octet $c\bar{c}$ pair in a ${}^1S_0$ or ${}^3P_J$ configuration, or to a color-singlet $c\bar{c}$ pair in a ${}^3S_1$ configuration.  Many previous studies \cite{Lee:2021oqr,Catani:2014qha,Ma:2014svb,Kang:2014tta,Sun:2012vc,Catani:2010pd,Mukherjee:2016cjw,Mukherjee:2015smo,Boer:2012bt,Echevarria:2019ynx,Fleming:2019pzj,DAlesio:2021yws,Boer:2020bbd,Bor:2022fga,Kishore:2021vsm,Scarpa:2019fol,DAlesio:2019qpk,Bacchetta:2018ivt,Mukherjee:2016qxa,Rajesh:2018qks,Godbole:2013bca,Godbole:2012bx,denDunnen:2014kjo,Kang:2014pya,Zhu:2013yxa} have investigated transverse momentum dependence in quarkonium direct production processes, but most, with the exception of Refs.~\cite{Echevarria:2020qjk,Copeland:2023wbu}, neglect TMD fragmentation.  Other papers \cite{vonKuk:2023jfd,Dai:2023rvd} have looked at TMD fragmentation functions for heavy quarks fragmenting to heavy hadrons in an effective field theory framework.

\begin{figure*}
    \includegraphics[trim=5.2cm 18.8cm 12cm 4.3cm,clip,scale=1.2]{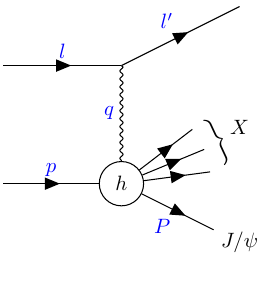}
    \caption{Semi-inclusive deep inelastic scattering.}
\label{fig: sidis}
\end{figure*}

In this paper we are interested in the production of quarkonium via light quark fragmentation so we study semi-inclusive deep inelastic scattering (SIDIS), shown in Fig.~\ref{fig: sidis}, at large $Q^2$ and small transverse momentum where this mechanism becomes relevant. We derive cross sections for the production of unpolarized, longitudinally polarized, and transversely polarized $J/\psi$ in SIDIS~\cite{Feynman:1969ej,Bloom:1969kc,Ji:2004wu} via fragmentation or photon-gluon fusion. The goal is to identify regions of phase space where one process clearly dominates, which would allow for a clean extraction of either the quark or the gluon TMDPDF. In addition, for regions where light quark fragmentation dominates, the $^3S_1^{[8]}$ long distance matrix element (LDME) gives the sole leading order contribution, implying that SIDIS at very large $Q^2$ offers a new avenue to extract this poorly constrained quantity. The kinematic regions we study are not accessible by current or previous experiments, like those conducted at JLab or HERA. Hence, this work is mostly exploratory in nature and meant to be compared directly with future experiments at the EIC.

We are not aiming for a precision calculation of cross sections and thus do not incorporate higher order corrections or resummations of large logarithms. Furthermore, we model the nonperturbative behavior of TMD functions with a Gaussian in transverse momentum, and the shape-function that arises in photon-gluon fusion to color octet $c\bar{c}$ as a Gaussian as well. The latter nonperturbative effects give contributions away from $z=1$, and have not been included in previous studies. The main novel features of this work are the comparison of TMD fragmentation with color octet production mechanisms and the study of polarized $J/\psi$ production in SIDIS at large $Q^2$. 
 
In Sec.~\ref{sec: sidis} we discuss the kinematics of SIDIS, the reference frame used in this calculation, and the NRQCD factorization formalism.  In Sec.~\ref{sec: pol} we review the treatment of the polarization of a spin-1 hadron such as the $J/\psi$.  In Secs.~\ref{sec: direct production} and \ref{sec: fragmentation} we detail the calculation of cross sections for $J/\psi$ production via photon-gluon fusion and light quark fragmentation, respectively.  In Sec.~\ref{sec: results} we plot the cross sections differential in $\Pv_\perp^2$ in different binned kinematic regions in order to isolate which production mechanisms are dominant in those regions. We also investigate the $\cos{(2\phi)}$ azimuthal asymmetry, which has been studied in detail in the literature \cite{Bor:2022fga,DAlesio:2021yws,Kishore:2021vsm,Boer:2020bbd,Scarpa:2019fol,DAlesio:2019qpk,Bacchetta:2018ivt,Mukherjee:2016qxa}. Including the effect of light quark fragmentation reduces this asymmetry. Finally, we conclude in section \ref{sec: conclusion}.

\section{Quarkonium production via SIDIS}
\label{sec: sidis}

\subsection{Kinematics}

We are considering $J/\psi$ production via SIDIS:
\begin{equation}
    \ell(l) + h(p) \rightarrow \ell(l^\prime)+H(P) + X \, ,
\end{equation}
where $\ell$ is a lepton, $h$ is the initial nucleon, $H$ is the final state $J/\psi$, whose momenta are indicated in parentheses. $X$ is the undetected part of the final state. We work in a frame where the nucleon and virtual photon momenta are back-to-back:
\begin{equation}
\begin{aligned}
    q^\mu = (0,0,0, -Q)\, ,\\
    p^\mu = \frac{Q}{2x_B}(1, 0, 0, 1)\,,
\end{aligned}
\label{eq: breit frame}
\end{equation}
and we define the usual kinematic variables \\
\begin{equation}
    Q^2 = -q^2 = -(l-l')^2, ~~x_B = \frac{Q^2}{2p\cdot q}, ~~y= \frac{p\cdot q}{p\cdot l}, ~~z = \frac{p\cdot P}{p\cdot q} \; .
\end{equation}
The transverse momentum of the $J/\psi$, $\Pv_\perp$, is with respect to the $z$-axis defined by the direction of the nucleon's momentum. The differential cross section for SIDIS is given by
\begin{equation}
    \frac{d\sigma}{dx \, dz \, dQ^2 \, d{\bf P}_\perp^2} = \frac{\alpha_{\rm em}^2 M}{2Q^2xzs} L^{\mu\nu}W_{\mu\nu} \, ,
\end{equation}
The leptonic and hadronic tensors are
\begin{equation}
    L_{\mu \nu} = e^{-2} \bra{l'} J_\mu(0)\ket{l}\bra{l} J^\dagger_\nu (0) \ket{l'} \, 
\end{equation}
\begin{equation}
W_{\mu \nu} = e^{-2}\int \frac{d^4 x} {(2\pi)^4} e^{-ixq} \sum_X \bra{p} J_\mu^\dagger(x) \ket{P,X}\bra{P,X} J_\nu(0) \ket{p} .
\end{equation}

In the parton model, one works with the initial and final state partons (such as quarks or gluons) instead of the hadrons. In this case we can define the momenta $p_A$ for the initial parton. The struck parton carries some fraction of the initial nucleon's longitudinal momentum, $\xi$, so new relevant variables can be defined 
\begin{equation}
    p_A^\mu = \xi p^\mu, ~~ x = \frac{x_B}{\xi} ~~ 
\end{equation}
which are more convenient to work with for factorization of the hadronic tensor. 

The component of the photon momentum transverse to $p_A$ and $P$ is
\begin{equation}
    q_t^\mu = q^\mu - \frac{1}{p_A\cdot P}\bigg(P\cdot q - \frac{p_A\cdot q}{p_A\cdot P}M^2\bigg)p_A^\mu - \frac{p_A\cdot q}{p_A\cdot P}P^\mu \, ,
\label{qt}
\end{equation}
such that $q_t \cdot q_t = -\qv_T^2$. Note that 
\begin{equation}
\begin{aligned}
P^0-P^3 =& \; \frac{2x}{Q} p_A\cdot P= \frac{2x}{Q}\xi p\cdot P = z Q \, ,\\
P^0+P^3 =& \; \frac{P_\perp^2+M^2}{P^0-P^3} = \frac{P_\perp^2+M^2}{zQ} \, ,
\end{aligned}
\end{equation}
so that 
\begin{equation}
    P^\mu = \frac{1}{2}\left(\frac{\Pv_\perp^2+M^2}{z Q} + zQ, 2{\bf P_\perp}, 0,\frac{\Pv_\perp^2+M^2}{z Q} - zQ\right) \, .
\end{equation}
Using the relation 
\begin{equation}
    \Pv_\perp = -z \qv_T,
\label{perp}
\end{equation}
where $\qv_T$ is the transverse momentum of the photon in a frame where the incoming proton and outgoing $J/\psi$ are collinear, the momentum of the $J/\psi$ is then given by
\begin{equation}
    P^\mu = \frac{Q z}{2}\left(\frac{\qv_T^2}{Q^2} + 1 +\frac{M^2}{z^2Q^2}, -\frac{2\qv_T}{Q}, 0, \frac{\qv_T^2}{Q^2} - 1 +\frac{M^2}{z^2Q^2}\right) \, .
\label{eq: breit frame P}
\end{equation}
The relation in Eq.~(\ref{perp}) is most easily understood if we write our momenta in light-cone coordinates $(q^+,q^-,\qv_T)$:
\begin{equation}
\begin{aligned}
    q^\mu =& \; \left(-\frac{Q}{\sqrt{2}},\frac{Q}{\sqrt{2}},{\bf 0} \right) \, , \\
    p_A^\mu =& \; \left(\frac{Q}{\sqrt{2} x},0,{\bf 0} \right) \, , \\
    P^\mu =& \; \left(\frac{\Pv_\perp^2 +M^2}{2P^-},P^-,\Pv_\perp \right) \, ,
\end{aligned}
\end{equation}
with $P^- = zQ/\sqrt{2}$.
Now consider the following Lorentz transformation \cite{Collins:2011zzd}
\begin{equation}
\begin{aligned}
    (V^+,V^-,{\bf V}_\perp) \rightarrow
    \left(V^+ +\frac{\qv_T^2}{Q^2} V^- + \frac{\sqrt{2}}{Q} \qv_T\cdot {\bf V}_\perp, V^-, {\bf V}_\perp + \qv_T \frac{\sqrt{2}}{Q} V^-\right) \, ,
\end{aligned}
\label{eq: boost}
\end{equation}
after the transformation $q^\mu =(q^+,q^-,\qv_T)$ with $q^-= Q/\sqrt{2}$, and 
\begin{equation}
\begin{aligned}
P_h^\mu =& \;  \left(\frac{\Pv_\perp^2 +M^2}{2P^-}+ \frac{\qv_T^2}{Q^2} P^-+\frac{\sqrt{2} }{Q} \qv_T \cdot{\bf P}_T,P^-,\Pv_\perp+\qv_T z \right) \, , \\
 =& \;  \left(\frac{M^2}{2 P^-},P^-, {\bf 0} \right) 
\end{aligned}
\end{equation}
where we set $\qv_T =-\Pv_T/z$, use $z=\sqrt{2}P^-/Q$, and denote the frame in which the $J/\psi$ has no transverse momentum with an $h$ subscript. Note that $p_A^\mu$ is unchanged by the Lorentz transformation. 
It is straightforward to use Eq.~(\ref{qt}) to show that $q_{t,h}= (0,0,\qv_T)$ in this reference frame.

\subsection{NRQCD factorization formalism}

We utilize the NRQCD factorization formalism, where the cross section for the production of a $J/\psi$ via SIDIS is written as a convolution between a TMD PDF of parton flavor $i$, $f_i$, with the short-distance coefficients $F_n$ associated with the production of a $c\bar{c}$ pair in spin and color state $^{2S+1}L_J^{[\rm color]}$ (collectively represented by $n$), times NRQCD long-distance matrix elements (LDMEs) $\braket{\mathcal{O}^{J/\psi}(n)}$, corresponding to the hadronization of the $c\bar{c}$ pair in state $n$ into a $J/\psi$ ,
\begin{equation}
    \begin{aligned}
        \sigma \sim \sum_{n,i} f_{i} \circledast F_{n, i} \braket{\mathcal{O}^{J/\psi}(n)} +{\cal O}\bigg(v, \frac{\Lambda}{p_\perp}\bigg).
    \end{aligned}
\label{eq: NRQCD fact}
\end{equation}
The utility of this formalism is that the LDMEs scale in the NRQCD power counting parameter $v$, the relative velocity of the quark/antiquark pair, which for charmonium goes as $v^2 \sim 0.3$. Thus Eq. (\ref{eq: NRQCD fact}) is a double expansion in $\alpha_s$ and $v$. For the scope of this paper, we need four LDMEs: $\braket{\mathcal{O}^{J/\psi}(^3S_1^{[1]})}$, $\braket{\mathcal{O}^{J/\psi}(^3S_1^{[8]})}$, $\braket{\mathcal{O}^{J/\psi}(^1S_0^{[8]})}$, and $\braket{\mathcal{O}^{J/\psi}(^3P_0^{[8]})}$. The first of these scales as $v^3$ and the rest scale as $v^7$, and we discuss in detail the power counting that justifies retaining only these terms as we consider each contribution.

\subsubsection{Polarization puzzle}

$J/\psi$ production has been measured in $e^+e^-$, $pp$, $p\bar{p}$, $ep$, $\gamma p$, and $\gamma\gamma$ collisions,  and compared with corresponding theoretical calculations in NLO QCD for both the color-singlet and color-octet mechanisms.  Historically, when fitting to NRQCD predictions, there have been discrepancies in the LDMEs between different fits.  References \cite{Butenschoen:2011yh,Butenschoen:2012qr} performed global fits to the world's data at NLO in NRQCD and find LDMEs that are consistent with the relative $v$ suppression of the color-octet mechanisms compared to the color-singlet mechanisms.  However, using these LDMEs predicts the polarization of the $J/\psi$ to be transverse to its momentum at high transverse momentum ($P_T$), which is in conflict with data indicating it has no polarization.  In light of this so-called polarization puzzle, other authors \cite{Chao:2012iv,Bodwin:2014gia} have performed fits to only the higher values of $P_T$ and obtained better agreement.  Both of these papers argue that NRQCD factorization is most valid at high $P_T$, and Ref.~\cite{Bodwin:2014gia} used Altarelli-Parisi evolution to resum logarithms of $P_T/M$.  We summarize the extracted LDMEs for these three papers in Table \ref{tab:LDMEs}.  Note that there is substantial disagreement between the two high $P_T$ fits on the value of the $^3S_1^{[8]}$ matrix element. Light quark fragmentation to $J/\psi$ is dominated by the $^3S_1^{[8]}$ 
mechanism, so the computation of polarized TMDFFs for $J/\psi$ production in this paper provides new observables from which to extract this LDME and may shed some light on this discrepancy.

\begin{table*}
\caption{NRQCD LDMEs for various fits.}
\begin{tabular}{l||r|r|r|r|}
\cline{2-5}
 & $\begin{aligned} \braket{\mathcal{O}^{J/\psi}(^3S_1^{[1]})} & \\ \times \; {\rm GeV}^3 & \end{aligned}$ & $\begin{aligned} \braket{\mathcal{O}^{J/\psi}(^3S_1^{[8]})} & \\ \times 10^{-2} \; {\rm GeV}^3 & \end{aligned}$ & $\begin{aligned} \braket{\mathcal{O}^{J/\psi}(^1S_0^{[8]})} & \\ \times 10^{-2} \; {\rm GeV}^3 & \end{aligned}$ & $\begin{aligned} \braket{\mathcal{O}^{J/\psi}(^3P_0^{[8]})}/m_c^2 & \\ \times 10^{-2} \; {\rm GeV}^3 & \end{aligned}$ \\ \hline \hline
\multicolumn{1}{|l||}{B \& K \cite{Butenschoen:2011yh,Butenschoen:2012qr}} & $1.32\pm 0.20$ & $0.224\pm 0.59$ & $4.97\pm0.44$ & $-0.72\pm 0.88$ \\ \hline
\multicolumn{1}{|l||}{Chao et al. \cite{Chao:2012iv}} & $1.16\pm 0.20$ & $0.30\pm 0.12$ & $8.9\pm 0.98$ & $0.56\pm 0.21$ \\ \hline
\multicolumn{1}{|l||}{Bodwin et al. \cite{Bodwin:2014gia}} & $1.32 \pm 0.20$ & $1.1\pm 1.0$ & $9.9\pm 2.2$ & $0.49\pm 0.44$ \\ \hline
\end{tabular}
\label{tab:LDMEs}
\end{table*}

\subsection{$J/\psi$ production mechanisms}

We are considering contributions to $J/\psi$ production from photon-gluon fusion and quark fragmentation. For photon-gluon fusion, the perturbative short distance coefficient, $F_n$, is obtained from a squared amplitude for the direct production diagram $\gamma^* + g \rightarrow c\bar{c}$.  We consider the diagrams at $\mathcal{O}(\alpha_s)$ and $\mathcal{O}(\alpha_s^2)$ (Fig.~\ref{fig: photon-gluon fusion}).  The former has a short-distance coefficient proportional to $\alpha_s(Q)$ and has leading  contributions from the color octet $^1S_0$ and $^3P_J$ channels, whose LDMEs scale as $v^7$.  The latter has a short-distance coefficient proportional to $\alpha^2_s(Q)$ and has a leading  contribution in the color singlet $^3S_1$ channel, whose LDME scales as $v^3$. For the large $Q$ accessible at the EIC $\alpha_s(Q) \sim 0.1$ while in the charmonium system $v^2 \sim 0.3$, thus the extra suppression of the color octet matrix elements by $v^4$ is partially offset by one less factor of $\alpha_s(Q)$ in the short distance coefficient. Therefore, the color octet and color singlet diagrams are roughly the same order in the double power counting.


Production via fragmentation is the leading power in $P_\perp^2/Q^2$ contribution to the cross section and thus should dominate the direct production contributions at large $Q^2$. However, for $J/\psi$ production via fragmentation in SIDIS we need to consider the mechanism for fragmentation into $c\bar{c}$. The hard coefficient in SIDIS production of $J/\psi$ determines the power of $\alpha_s(Q)$ by which individual contribution are suppressed; for quark initiated processes it is $\alpha_s^0(Q)$, while having a gluon initiated process costs a power of $\alpha_s(Q)$. The light quark involved in the quark initiated process can fragment to a $c\bar{c}$ which hadronizes to a $J/\psi$. At leading order this comes with an additional suppression of $\alpha_s(M)^2$ in the strong coupling and $v^7$ in the NRQCD scaling. All other fragmentation processes are further suppressed compared to quark fragmentation from quark initiated processes. How the leading fragmentation contribution compares to the direct production process is best determined by a comparison of these contributions in the cross section.

For production via fragmentation, TMD factorization allows the hadronic tensor to be written as
\begin{equation}
    W^{\mu \nu} \approx 2 z \int d^2\kv_T \, d^2\pv_T ~\delta^{(4)}(\pv_T - \kv_T + \qv_T ){\rm Tr} \left[\gamma^\mu \Phi(\pv_T, x) \gamma^{\nu} \Delta(\kv_T, z) \right] \, ,
\label{eq: Factorized W}
\end{equation}
where $\Phi(\pv_T, x) $ is the TMD PDF of the parton and $\Delta(\kv_T, z)$ is the TMD FF for the parton fragmenting into a $J/\psi$ \cite{Bacchetta:2000jk}.  (For details regarding a proof of factorization for fragmentation to a light hadron, refer to Refs. \cite{Collins:1981va,Collins:1984kg,Catani:1996yz,Echevarria:2011epo,Echevarria:2012js}).  Here, NRQCD factorization comes into play in the TMD FF, where we write
\begin{equation}
\begin{aligned}
    \Delta_{i\to J/\psi}(z, \kv_T; \mu)  = \sum_n d_{i\to c\bar{c}} (z, \kv_T; \mu) \braket{{\cal O}^{J/\psi}(n)} \, +{\cal O}\bigg(v, \frac{\Lambda}{p_\perp}\bigg),
\end{aligned}
\label{eq: NRQCD FT}
\end{equation}
with perturbative coefficients $d_{i\rightarrow c \bar{c}}$. For quark fragmentation at $\mathcal{O}(\alpha_s^2)$, three diagrams plus their mirrors contribute (Fig.~\ref{fig: QuarkFeynDiag}).  The only LDME involved at this order is color octet $^3S_1$.

The leading fragmentation contribution thus is ${\cal O}(\alpha_s^2\, v^7)$, which appears to be subleading to the leading direct production mechanisms. Naive power counting, however, is misleading for two reasons. First, fragmentation is the leading contribution to the cross section in the infinite $s$ limit, with the direct production contributions suppressed by powers of $M^2/s$. Second, the fragmentation process takes place at the scale $2m_c$, so that the short-distance coefficient is proportional to $\alpha(2m_c)\sim 0.3$. As a result a more nuanced analysis is needed to determine which mechanism dominates.

In Ref.~\cite{Echevarria:2020qjk}, the authors argue that in the region $(1-z)\sim 1$ and in the limit $Q^2 \gg M^2$, the relative contribution from color singlet photon-gluon fusion $d\sigma(\gamma^*g)$ and quark fragmentation $d\sigma(\gamma^* q)$ can be estimated to be
\begin{equation}
    \frac{d\sigma(\gamma^*g)}{d\sigma(\gamma^* q)} \sim \bigg( \frac{M}{Qv^2}\bigg)^2 \, ,
\end{equation}
which for charmonium is about $100 \; {\rm GeV}^2/Q^2$. This motivates us to study $J/\psi$ production for $Q> 10\,{\rm GeV}$ in SIDIS. These $Q$ values are much larger than ever measured at HERA~\cite{Naroska:2000nw,H1:1999ujo}. Reference~\cite{Echevarria:2020qjk} operates away from $z=1$ and so the $^1S_0$ and $^3P_J$ mechanisms are excluded from the plots in their paper. This is because in lowest order perturbation theory they are proportional to $\delta(1-z)$. An important point of this paper is that  higher-order shape function effects become important near $z=1$ \cite{Fleming:1997fq}, which allows these mechanisms to contribute in the high-$z$ regions.  

\section{$J/\psi$ polarization}
\label{sec: pol}

In our calculations for $J/\psi$ production via SIDIS, several nonrelativistic tensor structures appear. These are matched onto vacuum matrix elements of NRQCD operators, in terms of the NRQCD heavy quark and antiquark fields $\psi$ and $\chi$, and a projection operator $\mathcal{P}_{J/\psi(\lambda)}$ that projects out the $J/\psi$ state of helicity $\lambda$ \cite{Braaten:1996jt}. The matrix elements relevant for our calculation are:
\begin{equation}
    \begin{aligned}
        M^2 \eta^{\prime\dagger} \sigma_i \xi^\prime \xi^\dagger \sigma_j \eta &\leftrightarrow \braket{\chi^\dagger \sigma_i \psi \, \mathcal{P}_{J/\psi (\lambda)} \, \psi^\dagger \sigma_j  \chi}  \, , \\
        M^2 q^\prime_m q_n \eta^{\prime\dagger} \sigma_i T^a \xi^\prime \xi^\dagger \sigma_j T^a \eta &\leftrightarrow \braket{\chi^\dagger \sigma_i \big(-\frac{i}{2}\overleftrightarrow{{\bf D}}_m \big)T^a \psi \, \mathcal{P}_{J/\psi (\lambda)} \, \psi^\dagger \sigma_j \big(-\frac{i}{2}\overleftrightarrow{{\bf D}}_n \big) T^a \chi}  \, , \\
        M^2 \eta^{\prime\dagger} \sigma_i T^a \xi^\prime \xi^\dagger \sigma_j T^a \eta &\leftrightarrow \braket{\chi^\dagger \sigma_i T^a \psi \, \mathcal{P}_{J/\psi (\lambda)} \, \psi^\dagger \sigma_j T^a \chi}  \, , \\
        M^2 \eta^{\prime\dagger} T^a \xi^\prime \xi^\dagger T^a \eta &\leftrightarrow \braket{\chi^\dagger T^a \psi \, \mathcal{P}_{J/\psi (\lambda)} \, \psi^\dagger  T^a \chi}  \, . \\
    \end{aligned}
\end{equation}
These vacuum matrix elements give the NRQCD LDMEs:
\begin{equation}
    \begin{aligned}
        \braket{\chi^\dagger \sigma_i \psi \, \mathcal{P}_{J/\psi (\lambda)} \, \psi^\dagger \sigma_j  \chi} &= \frac{2M}{3}  \epsilon^*_{\lambda i} \epsilon_{j \lambda} \braket{\mathcal{O}^{J/\psi}(^3S_1^{[1]})} \, , \\
        \braket{\chi^\dagger \sigma_i \big(-\frac{i}{2}\overleftrightarrow{{\bf D}}_m \big)T^a \psi \, \mathcal{P}_{J/\psi (\lambda)} \, \psi^\dagger \sigma_j \big(-\frac{i}{2}\overleftrightarrow{{\bf D}}_n \big) T^a \chi} &= 2M \epsilon^*_{\lambda i} \epsilon_{j \lambda}  \delta_{mn} \braket{\mathcal{O}^{J/\psi}(^3P_0^{[8]})} \, , \\
        \braket{\chi^\dagger \sigma_i T^a \psi \, \mathcal{P}_{J/\psi (\lambda)} \, \psi^\dagger \sigma_j T^a \chi} &= \frac{2M}{3} \epsilon^*_{\lambda i} \epsilon_{j \lambda}  \braket{\mathcal{O}^{J/\psi}(^3S_1^{[8]})} \, , \\
        \braket{\chi^\dagger T^a \psi \, \mathcal{P}_{J/\psi (\lambda)} \, \psi^\dagger  T^a \chi} &= 2M \braket{\mathcal{O}^{J/\psi}(^1S_0^{[8]})} \, . \\
    \end{aligned}
\end{equation}
Note that spin-triplet matrix elements for the $J/\psi$ are proportional to $\epsilon^*_{\lambda i} \epsilon_{j \lambda}$ \cite{Braaten:1996jt} due to spin symmetry, where $\epsilon_{\lambda i}$ is the polarization vector of the $J/\psi$ and $\lambda$ denotes the helicity which can be $\lambda=+1$, $0$, and $-1$. The spin singlet operator is independent of polarization. 

We are interested in polarizations that are longitudinal and transverse to the $J/\psi$'s direction of motion.  The transverse polarizations can be obtained by subtracting the longitudinal polarization from the unpolarized case. It is straightforward to write down  the longitudinal polarization.  Requiring that its Cartesian portion be along the same direction as the $J/\psi$'s momentum, along with $\epsilon_L^2=-1$ and $\epsilon_L\cdot P=0$, fixes it to be
\begin{equation}
     \epsilon_L^\mu = \frac{1}{M}\left(|\Pv|,P^0 \hat{\Pv} \right) \, .
\end{equation}
For the unpolarized cross section, the sum over polarization yields the standard result,
\begin{equation}
    \sum_{\rm pol} \epsilon^{*\mu} \epsilon^\nu = -g^{\mu\nu} + \frac{P^\mu P^\nu}{M^2} \, .
\end{equation}

\section{Direct production}
\label{sec: direct production}

\begin{figure*}[t]
\centering
\begin{minipage}{0.5\textwidth}
\centering
\subfloat[]{\includegraphics[trim=5.2cm 20cm 12.4cm 4.4cm,clip,scale=1.2]{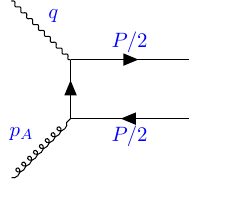}\label{octet}}
\end{minipage}%
\begin{minipage}{0.5\textwidth}
\centering
\subfloat[]{\includegraphics[trim=5.2cm 20cm 12.4cm 4.4cm,clip,scale=1.2]{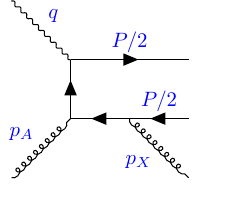}\label{singlet}}
\end{minipage}
\caption{Diagrams at (a) LO and (b) NLO in $\alpha_s$ contributing to quarkonium production via photon-gluon fusion. There are five other diagrams like the NLO graph, with permutations of the vertices.  Only the NLO graphs can produce a color singlet state.}
\label{fig: photon-gluon fusion}
\end{figure*}

At leading order in $\as$, out of the production mechanisms we consider, only the direct production of $J/\psi$ via photon-gluon fusion offers a probe of the gluon structure of the proton in SIDIS. At leading orders in $\alpha_s$ and the relative velocity $v$ the $c \bar{c}$ can be produced in either a color octet or color singlet state (shown in Fig.~\ref{fig: photon-gluon fusion}). The color octet contribution in Fig.~(\ref{octet}) contributes at $z=1$ in perturbation theory. It is important to study these production mechanisms because, in addition to offering access to the gluon content of the proton, they can compete with fragmentation in many kinematic regimes.

For direct production we employ a quasi-TMD factorization framework, working under the assumption that there is a regime where the transverse momentum of the PDF should be negligible and the majority of the transverse momentum dependence comes from the hadronization of the $c\bar{c}$ to the $J/\psi$. This is similar in practice to the collinear approach, with some slight modifications that are described below. 

For photon gluon fusion, the cross sections are quite lengthy and so we only give them in the $Q^2 \gg M^2, \, \Pv_\perp^2$ limit.  The full expressions, which are what we plot, can be found in a \emph{Mathematica} notebook on Github \cite{Copeland_JPsi_Production_NRQCD}, the link to which can also be found on this paper's arXiv posting.  We have checked that the expressions agree with the corresponding equations in Refs. \cite{Fleming:1997fq,Echevarria:2020qjk}.

\subsection{Color octet production}
\label{sec: octet pg fusion}

When $z\sim 1$ photon-gluon fusion predominantly produces $c \bar{c}$ in the color-octet state which is leading order in $\alpha_s$ (Fig.~\ref{octet}). In this limit, the $c\bar{c}$ absorbs all of the momentum of the collision, radiating no hard gluons. Hence $z = (p\cdot P)/(p\cdot q)\to 1$. To estimate the size of color-octet production we calculate this process within a collinear factorization framework. This has been computed before \cite{Fleming:1997fq}, and even been studied to some extent in the TMD framework \cite{Bacchetta:2018ivt, Boer:2020bbd}, so the results presented here are not new. At leading order the dominant channels are the $^1S_0^{[8]}$ and $^3P_{0}^{[8]}$ states and the final results are proportional to $\delta(1-z)$, confirming the basic arguments presented. The cross sections are:
\begin{equation}
    \begin{aligned}
        \frac{d\sigma_U(^1S_0^{[8]})}{dx \, dz \, dQ^2 \, d{\bf P}_\perp^2} \approx & \; \frac{32\pi^3 z \aem^2 \as}{9MQ^6\zt^3}\left(1-y+\frac{y^2}{2}\right)  \delta(\zb)\delta^{(2)}(\Pv_\perp)f_g(x\zt) \Oosz  \; ,
    \end{aligned}
\end{equation}
\begin{equation}
    \begin{aligned}
        \frac{d\sigma_L(^1S_0^{[8]})}{dx \, dz \, dQ^2 \, d{\bf P}_\perp^2} = \frac{1}{3} \frac{d\sigma_U(^1S_0^{[8]})}{dx \, dz \, dQ^2 \, d{\bf P}_\perp^2} \; , 
    \end{aligned}
\end{equation}
\begin{equation}
    \begin{aligned}
        \frac{d\sigma_U(^3P_0^{[8]})}{dx \, dz \, dQ^2 \, d{\bf P}_\perp^2} \approx& \; \frac{64\pi^3 z \aem^2 \as}{9M^3 Q^6 \zt^5}\left[ y^2 \left(8+z(3z-8) \right) +8(1-y)\zt -2(1-y)z\zt^2 \right] \\
        & \times \delta(\zb)\delta^{(2)}(\Pv_\perp)f_g(x\zt) \Otpz  \, ,
    \end{aligned}
\end{equation}
\begin{equation}
    \begin{aligned}
        \frac{d\sigma_L(^3P_0^{[8]})}{dx \, dz \, dQ^2 \, d{\bf P}_\perp^2} \approx& \; \frac{64\pi^3 z^3 \aem^2 \as}{9M^3 Q^6 \zt^5}\left[(2-y)^2 -2z(1-y) \right] \\
        & \times \delta(\zb)\delta^{(2)}(\Pv_\perp)f_g(x\zt) \Otpz \, ,
    \end{aligned}
\end{equation}
where $\zb = 1-z$, $\zt = 2-z$.
The above approximations are leading order in an expansion of small ${\bf P}_\perp$. We approximate this TMD function by replacing the $\delta^{(2)}(\Pv_\perp)$ with a Gaussian centered around $\Pv_\perp = {\bf 0}$. A similar argument applies to the $\delta(1-z)$, which is summed into a shape function which is peaked at $z=1$ and decays exponentially. Since we are interested in qualitative behavior of different contributions it is sufficient for our purposes to model the shape function with a Gaussian. A more thorough exploration of this process in a rigorous TMD framework should be conducted in future work. 

\subsection{Color singlet production}
\label{sec: singlet pg fusion}

When $ z < 1$ the color singlet channel dominates direct production, where one hard gluon is radiated away [Fig.~(\ref{singlet})]. There are six diagrams contributing to this process and the $J/\psi$ produced has transverse momentum dependence, even in the collinear limit. The cross sections for the color singlet mechanism are
\begin{equation}
    \begin{aligned}
        \frac{d\sigma_U(^3S_1^{[1]})}{dx \, dz \, dQ^2 \, d{\bf P}_\perp^2} \approx& \; \frac{512\pi \zb \aem^2 \as^2}{243M Q^6 z \zt^2(\Pv_\perp^2+\zb^2M^2)^2}\left(1-y+\frac{y^2}{2} \right) \\
        & \times \left[\Pv_\perp^2+\zb^2M^2(2-z\zt) \right] f_g(x) \Otsosing  \, , 
    \end{aligned}
\end{equation}
\begin{equation}
    \begin{aligned}
        \frac{d\sigma_L(^3S_1^{[1]})}{dx \, dz \, dQ^2 \, d{\bf P}_\perp^2} \approx& \; \frac{512\pi \zb \aem^2 \as^2}{243M Q^6 z \zt^2(\Pv_\perp^2+\zb^2M^2)^2}\left(1-y+\frac{y^2}{2} \right)\Pv_\perp^2 f_g(x) \Otsosing \, .
    \end{aligned}
\end{equation}

An interesting observation is that the longitudinal cross section vanishes at small $\Pv_\perp$ and the unpolarized cross section goes to a constant. This implies that the color singlet channel contributes only to transversely polarized $J/\psi$ for small $\Pp$. Also, the color singlet contribution is a subleading-power contribution in the TMD limit ({\it i.e.}~for small but nonzero $\Pv_\perp$ the color singlet contribution is down by $\Pv_\perp^2/M^2$ relative to the color octet contribution considered above).

\section{Quark fragmentation}
\label{sec: fragmentation}
\begin{figure*}[t]
\centering
\begin{minipage}{0.33\textwidth}
\centering
\includegraphics[trim=5.2cm 19.5cm 9.1cm 4.4cm,clip,scale=0.7]{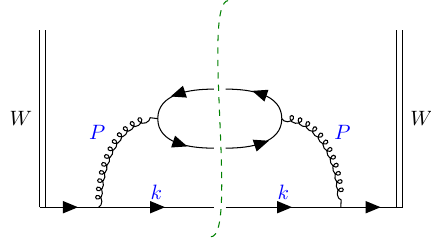}
\end{minipage}%
\begin{minipage}{0.33\textwidth}
\centering
\includegraphics[trim=5.2cm 19.5cm 9.1cm 4.4cm,clip,scale=0.7]{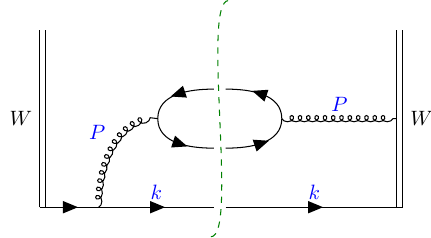}
\end{minipage}%
\begin{minipage}{0.33\textwidth}
\centering
\includegraphics[trim=5.2cm 19.5cm 9.1cm 4.4cm,clip,scale=0.7]{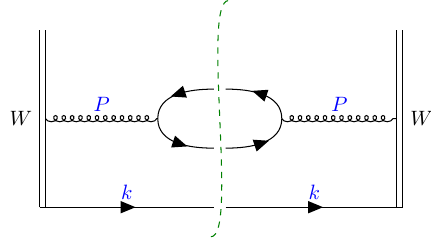}
\end{minipage}
\caption{Tree-level diagrams contributing to the polarized quark TMDFF.}
\label{fig: QuarkFeynDiag}
\end{figure*}

Fragmentation is a production process in which the virtual photon strikes a parton causing it to become highly energetic and fragment into final state hadrons. At leading twist, only light quark fragmentation is relevant for SIDIS; the diagrams for calculating the light quark fragmentation functions are depicted in Fig.~\ref{fig: QuarkFeynDiag}. The leading order polarized TMD FFs for a light quark fragmenting into $J/\psi$ were recently presented in Ref.~\cite{Copeland:2023wbu}. The results from this work are briefly reviewed below.

\subsection{TMDs}
The bare transverse momentum dependent fragmentation function for a quark of flavor $i$ to fragment into $J/\psi+X$, where $X$ represents other possible particles in the final state, is defined as

\begin{equation}
\begin{aligned}
\label{eq: q TMDFF}
\tilde{\Delta}^{(0)}_{q\to J/\psi} (z, \bv_T, P^+/z) =& \; \frac{1}{2 z N_c}{\rm Tr}\int \frac{d b^-}{2 \pi}e^{ib^-P^+/z}\sum_X \Gamma_{\alpha \alpha'} \\
& \times \bra{0} W_\lrcorner \psi_i^{ \alpha}(b) \ket{J/\psi (P) , X}  \bra{J/\psi (P), X} \bar{\psi}_i^{\alpha'} (b) W_\urcorner \ket{0}.
\end{aligned}
\end{equation}
Here the trace is over spin and color indices and $\Gamma \in \{\gamma^+/2, \gamma^+\gamma_5/2, i\sigma^{\beta +}\gamma_5/2\}$ covers the Dirac structures that project out the polarizations of the quark at leading twist.  The half staple shaped Wilson lines are defined as 
\begin{equation}
\begin{aligned}
    W_\lrcorner = &W_n ( b; + \infty , 0) W_{b_T}(+\infty n ; +\infty, b_T)\\
    W_\urcorner = &W_{b_T}(+\infty n ; +\infty, b_T) W_n ( 0 ; + \infty , 0) 
\end{aligned}
\end{equation}
with the usual Wilson lines, 
\begin{equation}
    W_n (x^\mu; a, b) =  {\cal P} \, {\rm exp}\left[i g \int_a^b ds~ n\cdot A^{a 0} (x + s n) t^a\right],
\end{equation}
where $n$ is a lightlike vector.


\subsection{Quark fragmentation functions}

In the case of an unpolarized parton and either unpolarized or longitudinally polarized $J/\psi$, the diagrams can be written in terms of the relevant fragmentation functions as:
\begin{equation}
    \begin{aligned}
        d_A + d_{B+\rm mirror} + d_c = D_1 + S_{LL} D_{1LL} \; ,
    \end{aligned}
\end{equation}
where $S_{LL}$ is a nonzero parameter in $\epsilon^*_\mu \epsilon_\nu$ \cite{Copeland:2023wbu}, and
\begin{equation}
    \begin{aligned}
        D_1 = & \; \frac{2\alpha_s^2}{27\pi z M^3} \frac{z^2 \kv_T^2(z^2-2z+2)+2M^2(z-1)^2}{[z^2\kv_T^2+M^2(1-z)]^2} \Otsooct \; , \\
        D_{1LL} =  & \; \frac{2\alpha_s^2}{27\pi z M^3} \frac{z^2 \kv_T^2(z^2-2z+2)-4M^2(z-1)^2}{[z^2\kv_T^2+M^2(1-z)]^2} \Otsooct \; . \\
    \end{aligned}
\end{equation}
The differential cross section is then
\begin{equation}
\begin{aligned}
\label{eq: fac cross}
    \frac{d \sigma_{UU}(l + H \to l' + J/\psi + X)}{dx ~dz ~dQ^2 ~d^2 {\bf P_\perp}} 
    = &\frac{4\pi \alpha^2}{Q^4} \left(1 - y +\frac{y^2}{2}\right) \bigg\{{\bf I}[f_1 D_1] + S_{LL} {\bf  I}[f_1 D_{1LL}]\bigg\} \, ,\\
\end{aligned}
\end{equation}
where the convolution integral $\bf I$ is
\begin{equation}
    \begin{aligned}
        {\bf I}[f~D] = \int d^2 \pv_T~d^2\kv_T~\delta^{(2)}(\pv_T-\kv_T+\Pv_\perp/z)~f(\pv_T)D(\kv_T) \, .
    \end{aligned}
\end{equation}
For unpolarized $J/\psi$, one then sums over $S_{LL} \in \{1/2,1/2,-1\}$, and for longitudinally polarized $J/\psi$ $S_{LL}\in\{-1\}$.  Note that summing over $S_{LL}$ cancels the cross section's dependence on $D_{1LL}$.

\section{Results} 
\label{sec: results}

\begin{figure}
    \centering
    \includegraphics[width = \linewidth]{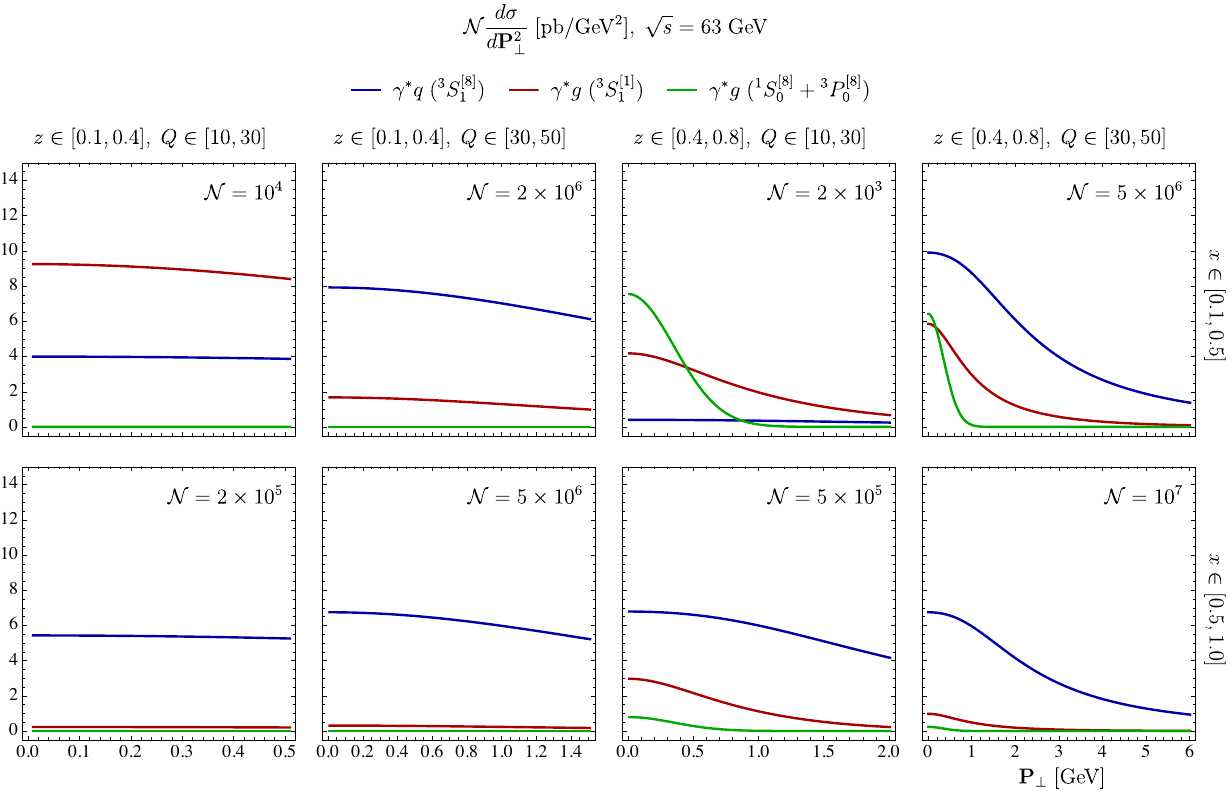}
    \caption{Binned contributions to the cross section for an unpolarized parton fragmenting to an unpolarized $J/\psi$, from direct production and light quark fragmentation.}
    \label{fig: cross sections 1}
\end{figure}
Using the factorized cross sections defined in Eq. (\ref{eq: fac cross}) and the collinear direct production cross sections, we can begin to make predictions for various processes. 

The delta functions in ${\bf P}_T$ and $(1-z)$ that appear in color octet photon gluon fusion are replaced by Gaussians (as discussed in \ref{sec: octet pg fusion}).
\begin{equation}
\begin{aligned}
   \delta^{(2)}({\bf P}_T) \to \frac{1}{\pi \left<P^2_T\right>}e^{-{{\bf P}_T^2}/\braket{P^2_T}}\\
   \delta(1-z) \to \frac{1}{\sqrt{\pi \braket{\bar{z}}} }e^{-{(1-z)^2}/\braket{\bar{z}}}.
\end{aligned}
\end{equation}
The parameters $\left<P^2_T\right>$ and $ \braket{\bar{z}}$ are chosen to be 0.25 $\rm GeV^2$ and 0.04, respectively. For this choice of $\braket{\bar{z}}$ the support of the shape function is between $z = 0.7 $ and $ 1$.

In the matching of the TMD FF, $J/\psi$ production occurs at a scale roughly around the mass of the $J/\psi$ so we evaluate the strong coupling at $\mu = 3.1$ GeV in the $J/\psi$ fragmentation functions. However, for direct production, the $J/\psi$ is produced at a much higher scale, so here we evaluate the strong coupling at $\mu = 30 $ GeV $\sim Q $. For both cases, the PDF is probed at the scale $\sim Q$ so we evaluate the PDFs at $\mu = 30$ GeV. In the numerical analysis we use the PDF sets for the up and down quarks from Ref.~\cite{Bastami:2018xqd}.

We plot the differential cross section at the center of mass energy $\sqrt{s} = 63$ GeV.  We divide the kinematic phase space into several bins of $x \in [0.1, 0.5] ~\&~ [0.5,1], z \in [0.1, 0.4] ~\&~ [0.4,0.8], $ and $Q $(GeV)$ \in [10, 30]~\&~ [30, 50]$, and plot in the TMD regime $\Pv_\perp \in [0, z_{\rm [bin~ min]} Q_{\rm [bin ~min]}/2]$, in the same manner as Ref.~\cite{Echevarria:2020qjk}. The results for an unpolarized beam scattering off an unpolarized cross section to produce unpolarized $J/\psi$ are presented in Fig.~\ref{fig: cross sections 1}

We notice in the region $x \in [0.1, 0.5]$, $Q \in [10,30]$ direct production dominates. The smaller $z$ region is largely populated by color singlet production and cleanly separates from direct color-octet production. However, in the larger $z\in [0.4, 0.8]$ bin the color octet mechanisms become relevant due to the implemented shape functions. To access the gluon TMD PDFs, it appears that the regions $x \in [0.1, 0.5]$, $Q \in [10,30]$, $z\in [0.1, 0.4]$ and $x \in [0.1, 0.5]$, $Q \in [10,30]$, $z\in [0.4, 0.8]$ may be ideal. 

When $x$ is larger, in the bin $[0.5, 1]$, fragmentation dominates over all other mechanisms, regardless of the $z$ and $Q$ regions considered. In fact, the fragmentation mechanism is essentially isolated in the bins where $x \in [0.5, 1]$ and $z \in [0.1, 0.4]$ meaning that this region is ideal for accessing information about the quark TMD PDFs. 

We have used the values for the NRQCD LDMEs determined by Chao {\it et~al.~}\cite{Chao:2012iv}:  $\braket{\mathcal{O}^{J/\psi}(^3S_1^{[8]})}$ = $0.3 \times 10^{-2}$ GeV$^3$, $\braket{\mathcal{O}^{J/\psi}(^3S_1^{[1]})}$ = $1.16$ GeV$^3$ , $\braket{\mathcal{O}^{J/\psi}(^1S_0^{[8]})}$ = $8.9 \times 10^{-2}$ GeV$^3$, and $\braket{\mathcal{O}^{J/\psi}(^3P_J^{[8]})}/m_c^2$ = $0.56 \times 10^{-2}$ GeV$^3$. Using the Bodwin {\it et~al.~}\cite{Bodwin:2014gia} values would increase the fragmentation contribution by a factor of $\sim 4$.  A goal of this work is to provide new observables and thereby be a first step toward a new global analysis of the world's data through which the NRQCD LDMEs can be more accurately extracted.  However, a more rigorous development of the production mechanisms in the TMD framework should be developed first before any precise extractions can be made.

\subsection{Polarized $J/\psi$ production}

\begin{figure}
    \centering
    \includegraphics[width = \linewidth]{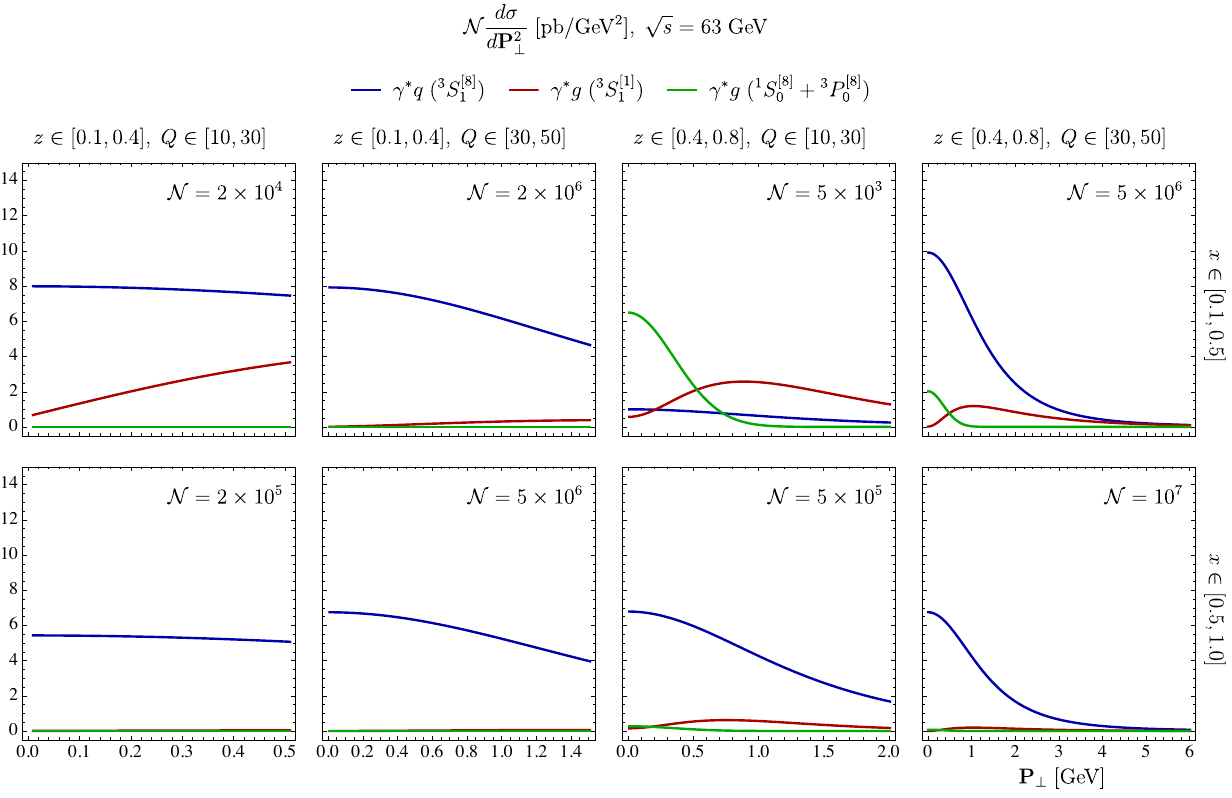}
    \caption{Binned contributions to the cross section for an unpolarized parton going to a longitudinally polarized $J/\psi$, from direct production and light quark fragmentation.}
    \label{fig: cross sections 2}
\end{figure}

We can study the production of polarized $J/\psi$ from unpolarized beams scattering off of unpolarized targets as well. In addition to giving more insight about where the TMDs may be extracted, studying polarized $J/\psi$ production provides an additional tool to isolate the various production mechanisms. This will be useful for determining the NRQCD LDMEs to greater precision in the future. The results for longitudinally polarized $J/\psi$ are presented in Fig.~\ref{fig: cross sections 2}. The main takeaway is that longitudinally polarized color singlet production is suppressed at low $\Pv_\perp$ and light quark fragmentation dominates for almost all bins, except for $x \in [0.1, 0.5]$, $Q \in [10,30]$, $z\in [0.4, 0.8]$. Transversely polarized $J/\psi$ (= unpolarized - longitudinal) would show the color singlet production becoming more prominent in both $x \in [0.1, 0.5]$, $Q \in [10,30]$ bins.

\begin{figure}
    \centering
    \includegraphics[width = \linewidth]{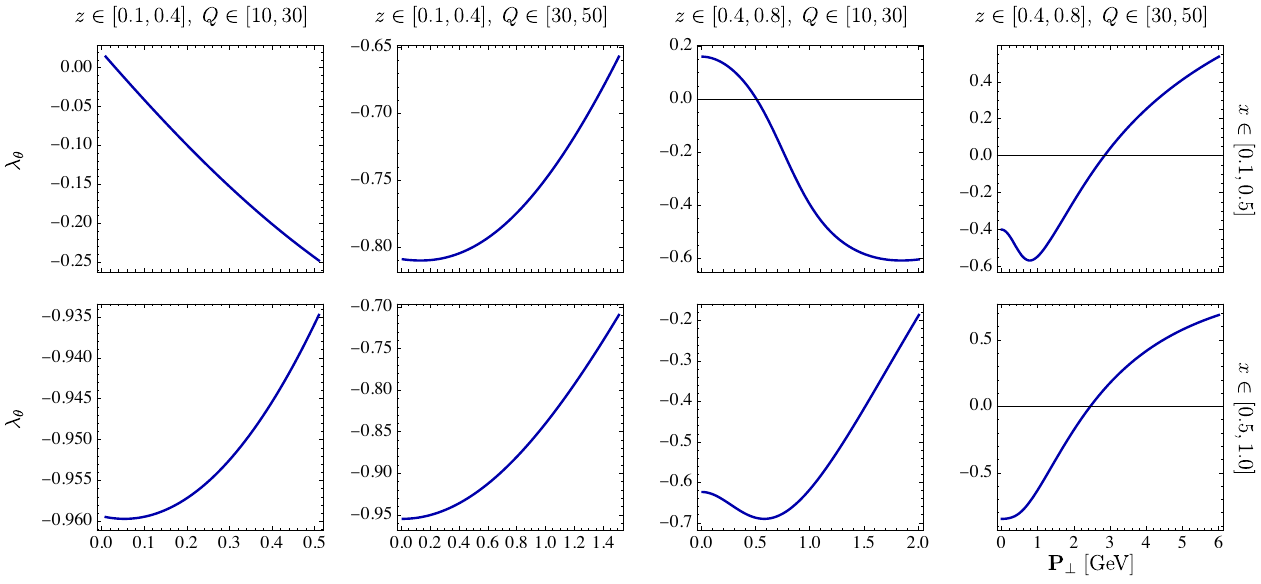}
    \caption{Sum over all contributions in each bin to the decay distribution parameter $\lambda_\theta$.}
    \label{fig: lambda theta}
\end{figure}

The polarization of the $J/\psi$ is measured via the angular distribution of the $J/\psi$'s decay to a $\ell^+\ell^-$ pair. The decay distribution can be parametrized as
\begin{equation}
    \frac{d\Gamma(J/\psi \rightarrow \ell^+\ell^-)}{d(\cos{\theta})} \propto 1 + \lambda_\theta \cos^2 \theta \; ,
\end{equation}
where the parameter $\lambda_\theta$ is
\begin{equation}
    \lambda_\theta = \frac{1-3\frac{\sigma_L}{\sigma_U}}{1+\frac{\sigma_L}{\sigma_U}}.
\end{equation}
We can plot the contributions to this parameter in each of the bins studied previously; the results are shown in Fig.~\ref{fig: lambda theta}. We see that $J/\psi$ is primarily longitudinally polarized in almost all bins.

\subsection{Azimuthal asymmetries}
\begin{figure}
    \centering
    \includegraphics[width = \linewidth]{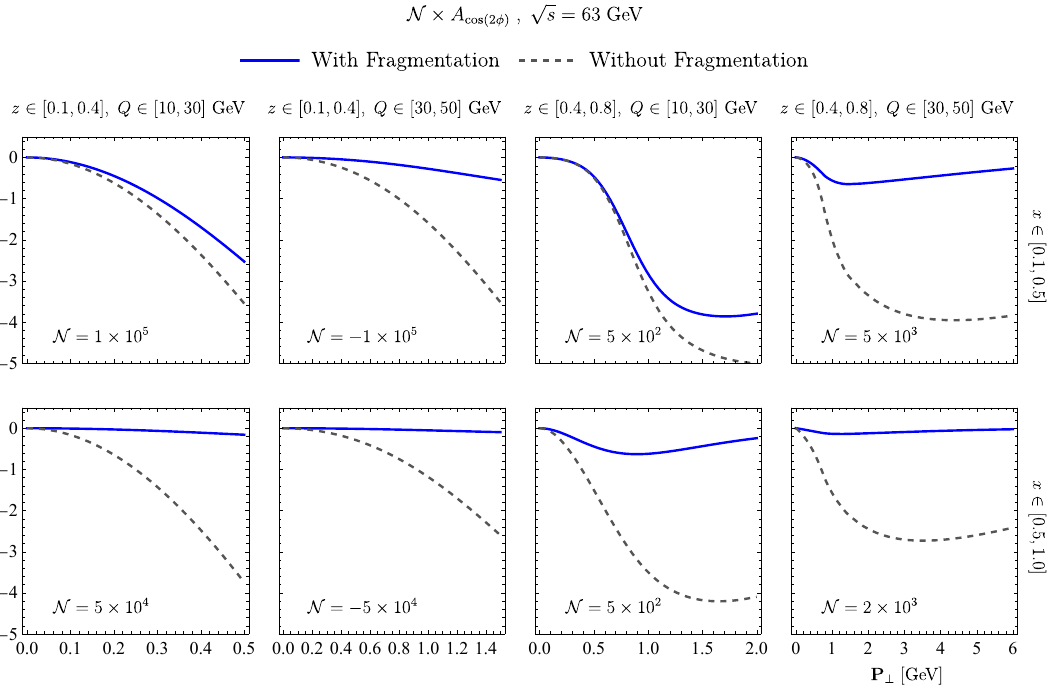}
    \caption{Binned contributions to the $\cos(2\phi_h)$ azimuthal asymmetry for an unpolarized parton producing an unpolarized $J/\psi$.}
    \label{fig: Asym}
\end{figure}

Azimuthal asymmetries are experimental observables of considerable interest to the community \cite{Bor:2022fga,DAlesio:2021yws,Kishore:2021vsm,Boer:2020bbd,Scarpa:2019fol,DAlesio:2019qpk,Bacchetta:2018ivt,Mukherjee:2016qxa}. They are defined as 
\begin{equation}
    A_W(\Pv_\perp) = \frac{\int d\phi_h \; W(\phi_h) d\sigma}{\int d\phi_h \; d\sigma}
\end{equation}
where $\phi_h$ is the angle between the $J/\psi$ production plane and the lepton plane and $W(\phi_h)$ is a weight dependent on the angle. In general, our cross sections have the form  
\begin{equation}
\begin{aligned}
    \frac{d\sigma}{dx dz dQ^2  d\phi_h d\Pv_\perp^2} = & \, [A_0(\Pv_\perp) + A_1( \Pv_\perp) \cos (\phi_h)+ A_2( \Pv_\perp) \cos(2\phi_h)]f_1^g(x) +  B_0(\Pv_\perp) f^q_{1}(x).
\end{aligned}
\end{equation}
The $x,z,$ and $Q^2$ dependence in the $A_i$ coefficients have been omitted for convenience. Since the quark fragmentation contribution has no azimuthal dependence it can be seen that azimuthal asymmetries provide a direct probe of the gluon PDFs in the proton. For example, it is common to study $A_{\cos(2\phi_h)}$ by plugging in $\cos(2\phi_h)$ for $W(\phi_h)$,
\begin{equation}
    A_{\cos(2\phi_h)}(\Pv_\perp) = \frac{\int dx \; A_2(\Pv_\perp) f_1^g(x)}{2\int dx \; [A_0(\Pv_\perp)f_1^g(x) +  B_0(\Pv_\perp) f^q_{1}(x)]}.
\end{equation}
The denominator is proportional to the unpolarized cross section which would be obtained by summing all of the contributions in Fig.~\ref{fig: cross sections 1}. The numerator offers direct access to the gluon PDF. 
Typically these asymmetries are only studied using the contributions from color singlet and/or color octet direct production. However, as shown in Fig.~\ref{fig: Asym}, including  fragmentation suppresses the predicted asymmetries dramatically regardless of the kinematic region. We do not expect this qualitative observation to change in a more rigorous application of the TMD framework (i.e. by taking the PDFs to have nontrivial TMD dependence etc.). 
Furthermore, we notice the asymmetry is by far the largest in the region $z\in [0.4, 0.8]$, $Q \in [10, 30]$ GeV and $x \in [0.1, 0.5]$. Thus, we suggest this kinematic region to probe the azimuthal asymmetry in $J/\psi$ production.
%
\section{Conclusion}
\label{sec: conclusion}

In this paper we compared SIDIS cross sections at large $Q^2$ for polarized $J/\psi$ production via photon-gluon fusion and light quark fragmentation utilizing an approximate TMD framework for the first time.  We identified kinematic regimes where photon-gluon fusion is dominant over fragmentation, which informs where gluon TMDs can be accessed. This is of high interest to the hadronic physics community. Likewise, there are many regions where quark fragmentation accounts for most of the $J/\psi$ produced. Thus, $J/\psi$ production in SIDIS at large $Q^2$ offers a new avenue for extracting the poorly constrained $^3S_1^{[8]}$ LDME.

In addition to the cross sections, we also investigated the decay distribution parameter $\lambda_\theta$, and the effect of including fragmentation in the azimuthal asymmetry $A_{\cos{(2\phi)}}$, which has not been included in previous studies of the asymmetry. In general, we find that including the effect of fragmentation suppresses these asymmetries, meaning this contribution is non-negligible. Our results will be valuable in comparing to experimental results at the upcoming EIC. 

Future research directions can encompass several areas. One such avenue is the exploration of higher-order corrections, including an examination of rapidity divergences. Additionally, there is potential for a full TMD treatment of the cross sections. Further investigations could focus on $J/\psi$ production in the context of $e^+e^-$ collisions. Lastly, an essential aspect of future work would involve the derivation of evolution equations for the TMD PDFs and TMD FFs appearing in these cross sections.

{\bf Acknowledgments} - We thank Alexey Prokudin for helpful discussions. M.~C., R.~H., and T.~M. are supported by the U.S. Department of Energy, Office of Science, Office of Nuclear Physics under grant Contract Nos.~DE-FG02-05ER41367. R.~H. and T.~M. are also supported by  the Topical Collaboration in Nuclear Theory on Heavy-Flavor Theory (HEFTY) for QCD Matter under award no.~DE-SC0023547. M.C. is supported by the National Science Foundation Graduate Research Fellowship under Grant No.~DGE 2139754. R.~G. and S.~F. are supported by the U.S. Department of Energy, Office of Science, Office of Nuclear Physics,  under award No.~DE-FG02-04ER41338. 


\bibliography{main}

\end{document}